\documentclass{appolb}
\usepackage{graphicx}
\usepackage{amssymb}
\usepackage{amsmath}
\usepackage{amsbsy}
\usepackage{lineno}
\begin{document}
\title{Beta-Decay Correlations in the LHC Era%
\thanks{Presented at Jagiellonian Symposium of Fundamental and Applied Subatomic Physics, Cracow, 2015; will be published in Acta Phys. Pol. B. 47, 349 (2016)}%
}
\author{Kazimierz Bodek
\address{The M. Smoluchowski Institute of Physics, Jagiellonian University, \\{\L}ojasiewicza 11, 30-348 Krak{\'o}w, Poland}
}
\maketitle
\begin{abstract}
Neutron and nuclear beta decay correlation coefficients are linearly sensitive to the exotic scalar and tensor interactions that are not included in the Standard Model. The proposed experiment will measure simultaneously 11 neutron correlation coefficients ($a$, $A$, $B$, $D$, $H$, $L$, $N$, $R$, $S$, $U$, $V$) where 5 of them ($H$, $L$, $S$, $U$, $V$) were never addressed before. Silicon pixel detectors are considered as promising alternative to multi-wire gas chambers devoted for electron tracking in the original setup. The expected sensitivity limits for $\epsilon_S$ and $\epsilon_T$ -- EFT parameters describing the scalar and tensor contributions to be extracted from the transverse electron polarization related coefficients $H$, $L$, $N$, $R$, $S$, $U$, $V$ are discussed.
\end{abstract}

\section{Introduction}
The observed excess of matter over antimatter in the Universe remains one of the most important puzzles in physics and unknown sources of fundamental CP symmetry violation are its necessary precondition. Intensive searches for new physics beyond the Standard Model (BSM) concentrate on two main frontiers: (i) high energy experiments performed at collider accelerators and (ii) low energy precision experiments. While the first group looks for exotic particles produced on-shell in high energy collisions, the second ones seek for tiny deviations in low energy observables which can be attributed to nonexisting in the SM exotic interactions. Nuclear and neutron beta decays are prominent examples of the second group. The persistent problem connected with this duality is to establish a common language in order to compare the results and allow for combined analyzes. Presently, the best solution is offered by the effective field theory (EFT) which can be safely applied to both experimental domains as long as the scale for exotic phenomena is sufficiently higher than the energy achievable in accelerators. In a number of seminal papers \cite{ref1,ref2,ref3,ref4,ref5} the EFT projection was applied to the charged current (CC) weak interaction including yet not observed but allowed by Lorentz invariance genuine scalar and tensor contributions. The EFT parametrization of measured observables in nuclear, neutron and pion decay and of the cross section for $pp \rightarrow e\bar{\nu} + ...$ process at high transverse mass led to the combined limits for exotic scalar and tensor interactions showing that, presently, the constraints obtained from both approaches are on a comparable accuracy level. Moreover, the EFT analysis describes prospects and defines challenges for future experiments in the low- and high-energy searches for BSM physics.

The above described thorough parametrization and analysis is, however, applied to the observables measured in the past and addressed in ongoing and prepared experiments. In this paper, we recall the idea of an entirely new experiment devoted to a simultaneous measurement of 11 correlation coefficients in the neutron decay where 5 of them were never attempted before \cite{ref6}. 7 of these correlation coefficients are related to the transverse polarization component of the electrons emitted in the decay. The proposed alternative offers a potential sensitivity comparable or better to that of ``classical'' correlation experiments and completely different systematics. As compared to Ref. \cite{ref6} new experimental options are discussed.

\section{Neutron $\beta$-decay correlation coefficients}
The transverse electron polarization is reflected in the distribution of the decay products via a number of the correlation coefficients relating it to other vectors characterizing the system, the most important being the electron and antineutrino momenta $\textbf{p}_e,\;\textbf{p}_{\bar{\nu}}$ and the neutron spin $\textbf{J}$. The corresponding formula limited to the lowest order terms can be found in the classical papers \cite{ref7,ref8,ref9}. Dropping out all the terms not depending explicitly on the transverse components of the electron polarization and retaining five exceptions: $a$, $b$, $A$, $B$ and $D$ (``classical'' correlations) one arrives at:
\begin{eqnarray}\label{EQN_JTW1}
\omega(E_e,\Omega_e,\Omega_{\bar{\nu}}) &\propto& 1 + a\,\frac{\textbf{p}_e\cdot\textbf{p}_{\bar{\nu}}}{E_eE_{\bar{\nu}}} + b\,\frac{m_e}{E_e} + \frac{\langle\textbf{J}\rangle}{J}\cdot\left[ A\,\frac{\textbf{p}_e}{E_e} + B\,\frac{\textbf{p}_{\bar{\nu}}}{E_{\bar{\nu}}} + D\,\frac{\textbf{p}_e\times\textbf{p}_{\bar{\nu}}}{E_eE_{\bar{\nu}}} \right] \nonumber \\
&+& \boldsymbol\sigma_\perp\cdot\left[H\,\frac{\textbf{p}_{\bar{\nu}}}{E_{\bar{\nu}}} + L\,\frac{\textbf{p}_e\times\textbf{p}_{\bar{\nu}}}{E_eE_{\bar{\nu}}} +  N\,\frac{\langle\textbf{J}\rangle}{J} + R\,\frac{\langle\textbf{J}\rangle\times\textbf{p}_e}{J\,E_e} \; + \right. \nonumber \\& &  \;\;\;\;\;\;\;\;\;\;\left. S\,\frac{\langle\textbf{J}\rangle}{J}\frac{\textbf{p}_e\cdot\textbf{p}_{\bar{\nu}}}{E_eE_{\bar{\nu}}} + U\,\textbf{p}_{\bar{\nu}}\frac{\langle\textbf{J}\rangle\cdot\textbf{p}_e}{J\,E_eE_{\bar{\nu}}} + V\,\frac{\textbf{p}_{\bar{\nu}}\times\langle\textbf{J}\rangle}{J\,E_{\bar{\nu}}} \right] ,
\end{eqnarray}
where $\boldsymbol\sigma_\perp$ represents a unit vector perpendicular to the electron momentum $\textbf{p}_e$ and $J=|\textbf{J}|$. In the infinite neutron mass approximation (no recoil), making usual assumptions for the Standard Model: $C_V=C'_V=1$ and $\lambda \equiv C_A=C'_A=-1.2701$ \cite{ref10}, and neglecting the contributions quadratic (and higher order) in $C_S,C'_S$, $C_T,C'_T$ one can express all the correlation coefficients from Eq.~(\ref{EQN_JTW1}) (called here $X$) as linear combinations of the real and imaginary parts of the scalar, $\mathfrak{S}$, and tensor, $\mathfrak{T}$, exotic couplings:
\begin{eqnarray}\label{EQN_JTW2}
X = X_{\mathrm{SM}} + X_{\mathrm{FSI}} + c_{\mathrm{Re}S}^X\Re(\mathfrak{S}) + c_{\mathrm{Re}T}^X\Re(\mathfrak{T}) + c_{\mathrm{Im}S}^X\Im(\mathfrak{S}) +c_{\mathrm{Im}T}^X\Im(\mathfrak{T})
\end{eqnarray}
with
\begin{eqnarray}
\mathfrak{S} \equiv \frac{C_S+C'_S}{C_V},\;\;\;\;\;\;\mathfrak{T} \equiv \frac{C_T+C'_T}{C_A}.
\end{eqnarray}
The coefficients $c$ in this expression are functions of $\lambda$ and kinematical quantities. Table~2 of Ref. \cite{ref6} summarizes their values calculated with the kinematical factors averaged over the electron spectrum in the kinetic energy range 200--783 keV. $X_{\mathrm{SM}}$ is the SM value of $X$ and the electromagnetic corrections called $X_{\mathrm{FSI}}$ were calculated in the static Coulomb field approximation with point-like proton and including only the contributions linear in the fine structure constant $\alpha$ \cite{ref8,ref9}.

\section{Scalar and tensor contributions to weak interaction in EFT approach}
\subsection{$\beta$-decay}
In order to bridge the above $\beta$-decay formalism with high-energy physics and permit sensitivity comparison of these processes with other low-energy charged-current observables and also with measurements carried out at high energy colliders, the model-independent EFT framework is employed. The effective nucleon-level couplings $C_i$, $C'_i$ ($i \in \{V, A, S, T\}$) can be generally expressed as linear combinations of the quark-level parameters $\epsilon_i$, $\tilde{\epsilon}_i$ ($i \in \{L, R, S, T\}$) \cite{ref4}. In one of the simplest scenarios one neglects the couplings to the right-handed neutrinos and adopts the axial-vector form factor $g_A^{exp} \approx g_A^{qcd} [1-2\Re(\epsilon_R)]+\mathcal{O}(\epsilon_i^2)$  value from experiments. This leads to the fact that the correlation coefficients are affected by the combinations of $\mathfrak{S} = 2 g_S \epsilon_S$ and $\mathfrak{T} = 8 g_T \epsilon_S$. The presence of exotic scalar and tensor interactions introduces two form factors: $g_S$ and $g_T$, needed to convert the measured nucleon-level quantities to the quark-level parameters.  Recent lattice QCD determinations give $g_S = 1.02 \pm 0.11$ and $g_T = 1.059 \pm 0.029$ (in the $\overline{MS}$ scheme and at the renormalization scale $\mu = 2$ GeV) \cite{ref5,ref5a}. Given the smallness of $\epsilon_S$, $\epsilon_T$, the poor precision of $g_S$, $g_T$ is still acceptable for the accuracy of present correlation experiments. The imaginary parts of the scalar and tensor couplings parametrize CP-violating contributions.

\subsection{High energy $pp \rightarrow e + \mathrm{MET} + ...$ channel}
The natural way to study the high energy BSM physics effects that can be compared with $\beta$-decay experiments is to search for electrons and missing transverse energy (MET) in $pp \rightarrow e + \textrm{MET} + ...$ channel since it has the same underlaying partonic process as in $\beta$-decay ($\bar{u}d \rightarrow e\bar{\nu}$). The cross section expressed in EFT approximation is a function of the same EFT parameters $\epsilon_i$, $\tilde{\epsilon}_i$ ($i \in \{L, R, S, T\}$) as seen in $\beta$-decay observables. The analysis of 20 fb$^{-1}$ CMS collaboration data collected at 8 TeV \cite{ref11,ref12} lead to
\begin{eqnarray}
  |\epsilon_S| < 5.8 \times 10^{-3}, \;\;\; |\epsilon_T| < 1.3 \times 10^{-3}
\end{eqnarray}
at 90\% C.L. The planned measurements at 14 TeV (50 fb$^{-1}$) will improve these limits by a factor of two.

\subsection{Transverse electron polarization in neutron $\beta$-decay}
The coefficients $H$ through $V$ relating the transverse electron polarization to $\textbf{p}_e$, $\textbf{p}_{\bar{\nu}}$ and $\textbf{J}$ have several interesting features. They vanish in the SM and reveal variable size of the FSI contributions, from very small to easily measurable in the present experiments. And, last but not least, the dependence on real and imaginary parts of the scalar and tensor couplings alternates exclusively from one correlation coefficient to another with varying linear combination coefficients. This feature allows one to deduce a complete set of constraints for $\mathfrak{S}$ and $\mathfrak{T}$ from the neutron decay alone. However, as pointed out in Refs. \cite{ref2,ref4}, there appears a serious technical problem leading to suppression or even to complete loss of linear sensitivity to the exotic couplings since, instead of $X$, effectively, measurements deliver $\tilde{X}$:
\begin{eqnarray}
  \tilde{X} &=& \frac{X}{1+b \frac{m}{E_e}}
  \;\approx\; X_{\mathrm{SM}} + X_{\mathrm{FSI}} \\
  &+& \left[c_{\mathrm{Re}S}^X-c_{\mathrm{Re}S}^b(X_{\mathrm{SM}} + X_{\mathrm{FSI}})\right]\Re(\mathfrak{S})
  + \left[c_{\mathrm{Re}T}^X-c_{\mathrm{Re}T}^b(X_{\mathrm{SM}} + X_{\mathrm{FSI}})\right]\Re(\mathfrak{T}) \nonumber \\
  &+& c_{\mathrm{Im}S}^X\Im(\mathfrak{S}) +c_{\mathrm{Im}T}^X\Im(\mathfrak{T}) \nonumber
\end{eqnarray}
Fortunately, the transverse electron polatization related coefficients are safe with that respect: the cancellation term is proportional to $(X_{\mathrm{SM}} + X_{\mathrm{FSI}})$ which even in the extreme cases of large electromagnetic contribution ($H$, $N$) does not exceed 0.07 so that the potential loss of the linear sensitivity to the real part of the scalar and tensor couplings is significantly suppressed. This feature makes the transverse electron polarization related coefficients attractive for future searches of BSM physics.

\section{Experiment: the BRAND project}
The proposed in Ref. \cite{ref6} experiment (identified meanwhile with the acronym BRAND) will measure 11 correlation coefficients ($a$, $A$, $B$, $D$, $H$, $L$, $N$, $R$, $S$, $U$, $V$). The transverse electron polarization will be measured using the Mott scattering from a high $Z$ target (e.g. Pb) similarly as in the experiment of Refs. \cite{ref13,ref14}. The measurement of momenta of electrons and protons in coincidence allows to reconstruct the antineutrino momentum and accesses the terms dependent on this quantity. The analysis strategy is to reconstruct event-by-event the decay kinematics together with the decay origin and, in this way, reduce the effects due to the large size of the decay source. In particular, fixing the three-body decay kinematics by the measured electron energy and relative $e-p$ angle one realizes that the proton energy and thus the proton time-of-flight must choose between two discrete (and known) values.

\begin{figure}[htb]
\begin{center}
\includegraphics[width=12.5cm]{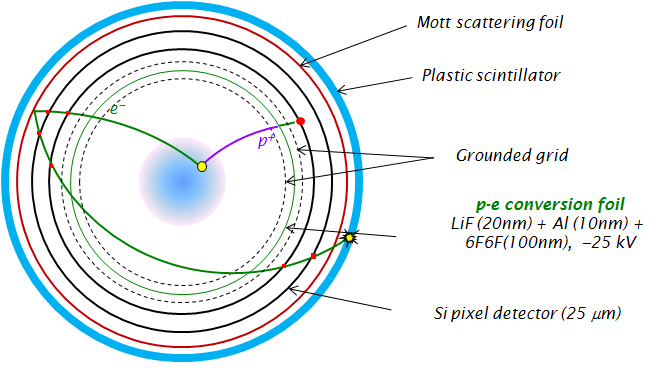}
\caption{\label{FIG1} (Color on-line) Sketch of the proposed experimental setup in the cylindrical geometry. Only the cross section perpendicular to the detector axis is shown.}
\end{center}
\end{figure}

The key ideas of the originally proposed setup were: (i) efficient cylindrical detector geometry, (ii) electron tracking in a multi-wire drift chamber (MWDC) with both wire ends readout, (iii) detection of both direct and Mott-scattered electrons in a plastic scintillator hodoscope, (iv) conversion of protons (accelerated to 20-30 keV) into bunches of electrons ejected from a thin LiF layer and (v) acceleration and subsequent detection of ejected electrons in a multi-wire proportional chamber (MWPC) with both wire ends readout. Such a setup is challenging for several reasons the most important being the implementation in vacuum of large area and thin window gas detectors. Meanwhile, it appeared an attractive alternative option for electron tracking which, potentially, can significantly simplify the experiment. It makes use of the impressive progress in technology of silicon pixel detectors, especially in so called MAPS (Monolithic Active Pixel Sensors) \cite{ref15}. Such detectors are produced in commercially available HV-CMOS process and can be as thin as 25 $\mu$m \cite{ref16}. Suitably configured MAPS can replace clumsy MWDC/MWPC and easily operate in vacuum. Their superior position resolution, high rate tolerance and very good fill factor (active-to-total area ratio) allows for compact geometry which will significantly relax experimental condition. A sketch of the experimental setup cross section is shown in Fig. \ref{FIG1}.

\begin{figure}[htb]
\begin{center}
\includegraphics[width=6.2cm]{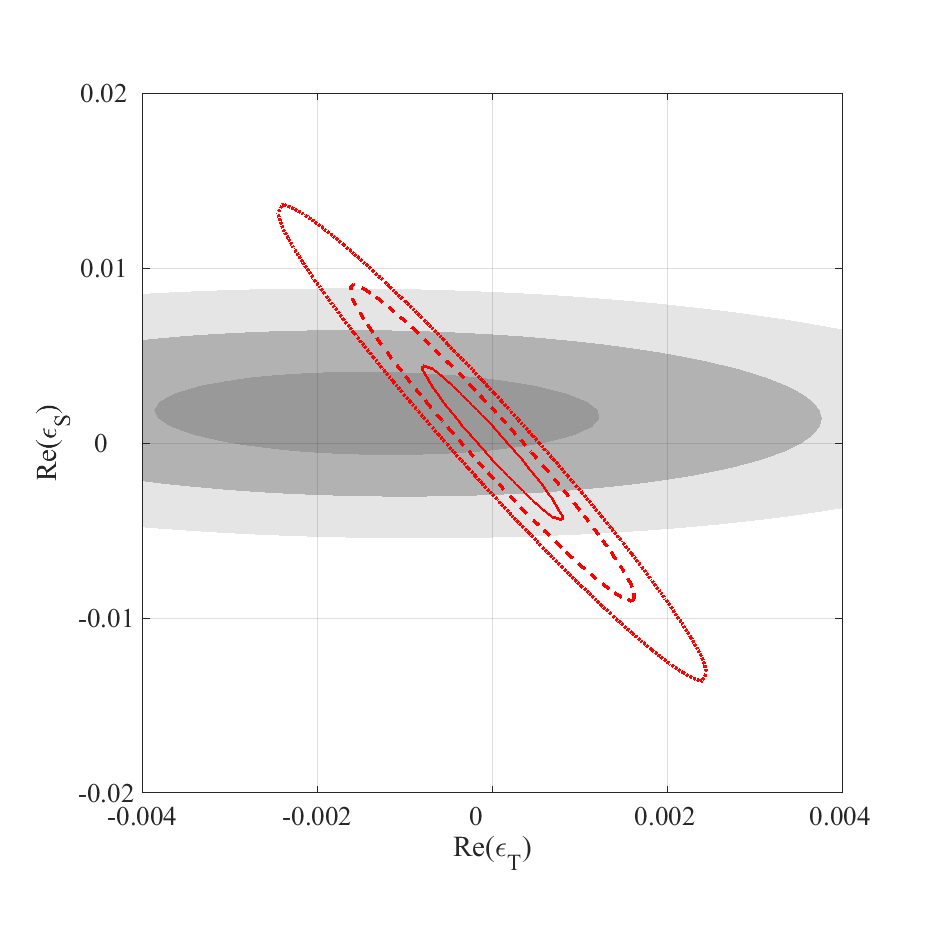}
\includegraphics[width=6.2cm]{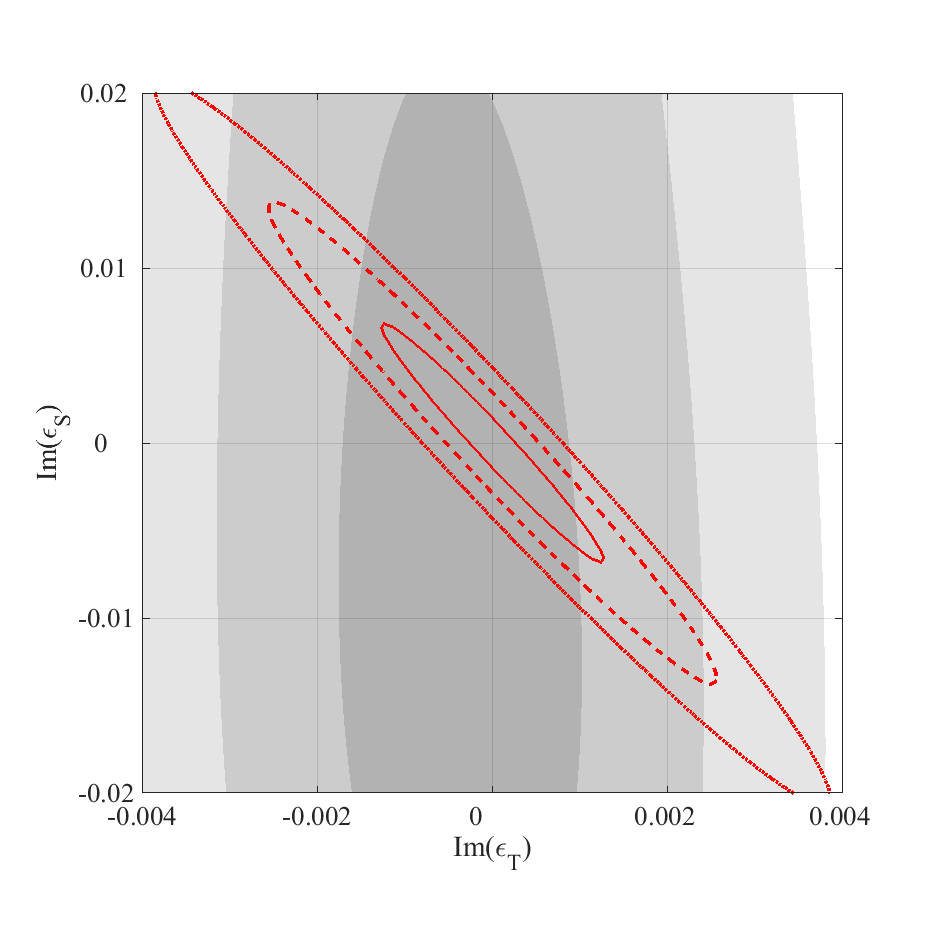}
\caption{\label{FIG2} (Color on-line) Experimental bounds on the scalar vs. tensor EFT couplings. The gray areas represent the information deduced from presently available experiments, while the red lines represent the limits resulting from the correlation coefficients $H$, $L$, $N$, $R$, $S$, $U$ and $V$ measured with the anticipated accuracy of $5\times 10^{-4}$. Solid, dashed and dotted lines correspond to 1-, 2- and 3- sigma confidence levels, respectively, in analogy to decreasing intensity of the grey areas.}
\end{center}
\end{figure}

The proposed solutions imply an unbiased registration of direct electrons (without Mott scattering) and also without accompanying protons. The experiment can be run in two modes: (1) unpolarized neutrons -- measurement of $a$, $H$, $L$ and (2) polarized neutrons -- measurement of $a$, $A$, $B$, $D$, $H$, $L$, $N$, $R$, $S$, $U$ and $V$ coefficients. It has been estimated that with $10^5$ decays per second in the fiducial volume and six month long data taking period one can achieve the anticipated sensitivity of about $5\times 10^{-4}$ for the transverse electron polarization related correlation coefficients. This, in turn, leads to significantly tighter bounds for the exotic coupling constants. In Fig. \ref{FIG2}, these bounds are compared with those extracted from existing experimental $\beta$-decay correlation coefficients.

It should be noticed that the BRAND experiment is not dedicated to the measurement of ``classical'' coefficients $a$, $A$, $B$, and $D$, however, their presence in the data will allow for the direct comparison of entirely different systematic effects in BRAND and in ``classical'' neutron decay correlation experiments.

\section{Conclusions}
Precise measurements of $\beta$-decay correlation coefficients compete vigorously with high energy experiments such as e.g. CMS at LHC in searches for exotic phenomena beyond the Standard Model. Emerging new generation of the neutron decay correlation experiments with the anticipated accuracy of $10^{-3} \div 10^{-4}$ will possibly outperform the ongoing experiments at high energy frontier. In this paper, we argue that the neutron decay correlation experiments can be significantly extended with new observables which are related to the transverse polarization of emitted electrons. They have several attractive features the most important being the immunity to cancelation of their linear sensitivity to the exotic couplings due to the Fierz term $b$ present in all experimentally extracted quantities. Nevertheless, it should be clear that the presented formulas are sufficient only for estimation of size of the expected effects. For rigorous extraction of the exotic coupling constants with the accuracy of $10^{-3} \div 10^{-4}$ the adequate formalism must include the recoil and radiative corrections which size is comparable to the anticipated accuracy.


\end{document}